%% file: Vega-preprint.tex
\documentclass{nature-SI}
\usepackage[dvips]{graphicx}

\newcommand{\etal}{et al.}

\newcommand{\kms}{\ensuremath{\mbox{km\,s}^{-1}}}

\newcommand{\Msun}{\ensuremath{\mbox{$M_{\odot}$}}}
\newcommand{\Rsun}{\ensuremath{\mbox{$R_{\odot}$}}}
\newcommand{\Lsun}{\ensuremath{\mbox{$L_{\odot}$}}}
\newcommand{\bea}{\begin{eqnarray*}}
\newcommand{\ba}{\begin{array}}
\newcommand{\bdm}{\begin{displaymath}}
\newcommand{\eea}{\end{eqnarray*}}
\newcommand{\ea}{\end{array}}
\newcommand{\edm}{\end{displaymath}}

\title{Vega is a rapidly rotating star}

\author{D.~M.\ Peterson$^1$, C.~A.\ Hummel$^{2,3}$, T.~A.\ Pauls$^4$, 
J.~T.\ Armstrong$^4$, J.~A.\ Benson$^5$, G.~C.\ Gilbreath$^4$, 
R.~B.\ Hindsley$^4$, D.~J.\ Hutter$^5$, K.~J.\ Johnston$^3$, 
D.\ Mozurkewich$^6$ \& H.~R.\ Schmitt$^{4,7}$}

\begin{document}

\maketitle

\begin{affiliations}
 \item Department of Physics and Astronomy, Stony Brook University, 
Stony Brook, New York, USA 11794-3800
 \item European Southern Observatory (ESO), Casilla 19001, Santiago 19,
 Chile
 \item US Naval Observatory, 3450 Massachusetts Avenue~NW, Washington, 
DC, USA, 20392-5420
 \item Naval Research Laboratory, Code 7215, 4555 Overlook Avenue~SW, 
Washington, DC, USA 20375 
 \item US Naval Observatory, Flagstaff Station, 10391 West Naval 
Observatory Road, Flagstaff, Arizona, USA 86001-8521
 \item Seabrook Engineering, 9310 Dubarry Road, Seabrook, Maryland, USA 20706
 \item Interferometrics, Inc., 13454 Sunrise Valley Drive, Suite 240, 
Herndon, Virginia, USA 20171 
\end{affiliations}

\input{First-paragraph}

\input{Text-nature}

\vspace{20pt}
\bibliography{RotatingStars-nature}
%\begin{thebibliography}{1}
%\bibitem{dummy} Articles are restricted to 50 references, Letters
%to 30.
%\bibitem{dummyb} No compound references -- only one source per
%reference.
%\end{thebibliography}

%% Here is the endmatter stuff: Supplementary Info, etc.
%% Use \item's to separate, default label is "Acknowledgements"

\begin{addendum}
 \item[Supplementary Information]is linked to the online version of the
paper at www.nature.com/nature.
 
 \item[Acknowledgements]We thank R. Kurucz for advice on calculating
 broadband fluxes using model atmospheres.  The NPOI facility is a
 collaboration between the Naval Research Laboratory and the US Naval
 Observatory in association with Lowell Observatory, and is funded by
 the Office of Naval Research and the Oceanographer of the Navy.  This
 research has made use of the SIMBAD literature database, operated at
 CDS, Strasbourg, France, and of NASA's Astrophysics Data System.

 \item[Correspondence] The authors declare that they have no competing
 financial interests.  Correspondence and requests for materials should
 be addressed to D.\,M.\,P.\ (email: dpeterson@astro.sunysb.edu).
\end{addendum}

%%
%% TABLES
%%
%% If there are any tables, put them here.
%%

\spacing{1}
\begin{table}
{\bfseries Table 1 \vline\ Vega Model and Derived Parameters
\label{tab:numbers}} 

%\begin{tabular}[b]{l@{\extracolsep{0cm}}c@{\extracolsep{0cm}}r
%@{\extracolsep{0cm}}r}\hline\hline

%\renewcommand{\arraystretch}{0.6}
\begin{tabular}[b]{l@{\extracolsep{4.92cm}}c@{\extracolsep{4.92cm}}c}\hline

Quantity & Value & Error (s.\ d.)*\\\hline

%\multicolumn{3}{c}{Model Parameters Fit}\\
$\omega=\Omega/\Omega_{\mathrm{B}}$\dag & 0.926  & $\pm$0.021 \\
$\theta_{\mathrm{p}}$ (mas)\dag        & 2.767  &    0.037 \\
$T_{\mathrm{p}}$ (K)\dag               & 9,988   &    61 \\
$i$ (deg.)\dag                          & 4.54   &    0.33 \\
$PA$ (deg.)\dag                         & 8.6    &    2.7 \\
%\multicolumn{4}{c}{Model Parameters Derived}\\
$v_{\mathrm{eq}}$ (\kms) & 274 & 14\\
$v_{\mathrm{eq,B}}\mbox{\,\ddag}$ (\kms) & 356.1 & 2.4\\
$\Omega$ (d$^{-1}$) & 1.884 & 0.081\\
$\Omega_{\mathrm{B}}\mbox{\,\ddag}$ (d$^{-1}$) & 2.034 & 0.041\\
$T_{\mathrm{eq}}$ (K) & 7,557 & 261\\
$R_{\mathrm{p}}$ (\Rsun) & 2.306 & 0.031\\
$R_{\mathrm{eq}}$ (\Rsun) & 2.873 & 0.026\\
$\theta_{\mathrm{min}}\S$ (mas) & 3.441 & 0.031\\
$\theta_{\mathrm{max}}\S$ (mas) & 3.446 & 0.031\\
$\log L$ (\Lsun) & 1.544 & 0.018\\
$\log g_{\mathrm{p}}$ (cm$^2$\,s$^{-2}$) & 4.074 & 0.012\\
$\log g_{\mathrm{eq}}$ (cm$^2$\,s$^{-2}$) & 3.589 & 0.056\\
%\multicolumn{4}{c}{Non-rotating Parameters\,\mbox{\ddag}}\\
$M\|$ (\Msun) & 2.303 & 0.024\\
%$\log g$ (cm$^2\,s$^{-2}$) & 4.066 & 0.013\\
$T_{\mathrm{eff}}\|$ (K) & 9,306 & 86\\
Age$\|$ (Myr) & 386 & 16\\\hline
\end{tabular} 
%\end{table}

%\begin{table}
%\renewcommand{\arraystretch}{0.6}
Six quantities are needed to uniquely define the Roche model of a
star\cite{Peterson06}: the ratio of the angular rotation to that of
breakup, $\omega = \Omega/\Omega_B$, the inclination (or tilt) of the
rotational axis to the line of sight, $i$, the position angle, $PA$, of
the pole on the sky, the radius of the polar axis, $R_{\mathrm{p}}$ (or
equivalently, using the parallax, the polar angular diameter,
$\theta_{\mathrm{p}}$), the effective temperature at the pole,
$T_{\mathrm{p}}$ and the surface gravity at the pole, $g_{\mathrm{p}}$
or equivalently, the mass.  It is then possible to calculate the
radius, $R(\phi)$, of the star for a given stellar latitude, $\phi$,
the effective gravity and the local temperature ($T(\phi)$).  For the
spectral calculations we have adopted the ATLAS model
atmospheres\cite{Kurucz93} and in particular the Van Hamme\cite{VH93}
limb-darkening parameterization of that grid.  Other parameters include
linear rotational velocities, $v$, the luminosity, $L$, surface
gravities, $g$, the mass, $M$, and effective temperature,
$T_{\mathrm{eff}}$, the last referring to the entire non-rotating star.
Subscripts "eq" and "p" specify quantities evaluated at the equator and
pole, respectively.\\
\mbox{*}Uncertainties due to the parallax have not been included in the
errors.\\
\mbox{\dag}The parameters derived from the model fit.  A mass of
2.30\,\Msun was assumed in the fit.\\
\mbox{\ddag}Rotating at breakup but with the same mass and polar radius\\ 
\mbox{\S}$\theta_{min,max}$ are the minimum and maximum projected
angular diameters.\\
\mbox{$\|$}The parameters of a non-rotating star from the Padova 
grid\cite{Girardi00} which would reproduce the (corrected)
luminosity and polar radius\cite{Peterson06}\\ 
\end{table}
%\vfill\eject

\begin{table}
{\bfseries Table 2 \vline\ Predicted Photometric ``Excesses'' for Vega
\label{tab:magnitudes}}

\begin{tabular}[b]{l@{\extracolsep{4.67cm}}c@{\extracolsep{4.67cm}}c}\hline
Band & Wavelength ($\mu$m) & Excess (mag)\\\hline
R & 0.67 & -0.016 \\
I & 0.86 & -0.029 \\
J & 1.12 & -0.052 \\
K & 2.14 & -0.072 \\
L & 3.69 & -0.072\\\hline
\end{tabular}

The differential excesses for the model of Vega compared to a
non-rotating model (both solar composition) which matches the $V$
magnitude with an angular diameter of 3.24\,mas (ref. 27).  Until the
question of composition is resolved, these values should be considered
indicative.
\end{table}
%\clearpage

\end{document}

%% file: First-paragraph.tex
{\bf Vega, the second brightest star in the northern hemisphere, serves
as a primary spectral type standard\cite{MK73}.  While its spectrum is
dominated by broad hydrogen lines, the narrower lines of the heavy
elements suggested slow to moderate rotation, giving confidence that
the ground-based calibration of its visibile spectrum could be safely
extrapolated into the ultraviolet and near-infrared (through atmosphere
models\cite{Bohlin04}), where it also serves as the primary photometric
calibrator.  But there have been problems: the star is too bright
compared to its peers\cite{Petrie64} and it has unusually shaped
absorption line profiles, leading some\cite{Gray88, Gulliver94} to
suggest that it is a distorted, rapidly rotating star seen pole-on.
Here we report optical interferometric observations of Vega which
detect the asymmetric brightness distribution of the bright, slightly
offset polar axis of a star rotating at 93\% of breakup speed.  In
addition to explaining the unusual brightness and line shape
pecularities, this result leads to the prediction of an excess of
near-infrared emission compared to the visible, in agreement with
observations\cite{Blackwell83, Leggett86b}.  The large temperature
differences predicted across its surface call into question composition
determinations, adding uncertainty to Vega's age and opening the
possibility that its debris disk\cite{IRAS} could be substantially
older than previously thought\cite{Habing99, Song01}}.

%% file: Text-nature.tex
Single baseline Michelson stellar interferometers measure complex
``visibilities''\cite{TMS94}, usually recorded as amplitudes and
phases, which are related to the intensity distribution of the target
through a Fourier Transform.  Even though the phases have been shown to
be very sensitive to asymmetries in the intensity distribution, they
are badly corrupted by the atmosphere and have been little used in the
optical.  But for closure phases, the data we focus on here--which are
obtained by summing the phases measured on each baseline of a triangle
in an interferometric array such as the Navy Prototype Optical
Interferometer\cite{NPOI} (NPOI)--the atmospheric contribution cancels.
The use of closure phase in the radio\cite{Jen58} has enabled a
dramatic gain in dynamic range of interferometric images made from
multi-antenna arrays.  In the optical\cite{COAST, NPOI}, where the
phase errors can reach 100 waves on long baselines, the technique
enables the use of phase information in any guise.  As the observations
reported here were made just with a three telescope array, the
application of imaging techniques was not justified and we relied
instead on fitting Roche models to the closure phase data.

The application of Roche spheroids to rotating stars was worked out 80
years ago\cite{vZ24}.  Assuming solid body rotation and a point mass
gravitational potential, a rotating star will adopt the figure of a
Roche spheroid.  Conservation of energy through surfaces of constant
potential leads to the prediction that when the energy is transported
by radiation the amount transported will vary over the surface in
proportion to the effective gravity\cite{HS68} (the net of gravity less
local centrifugal terms).  Near breakup, the effective gravity near the
equator can become quite small, leading to the prediction of a large
drop in the local temperature with a corresponding decrease in
brightness, an effect referred to as ``gravity darkening''.  Rapidly
rotating stars seen at intermediate inclinations are therefore expected
to display asymmetric intensity distributions.
Altair\cite{Peterson06,Dom05} proved to be the first major test of this
theory in an isolated rotating star, where the theory succeeded to a
high degree in describing a very non-trivial brightness distribution.

The observations considered here were obtained during late May and
early June 2001 on the same nights as those of Altair previously
reported\cite{Peterson06,Ohishi04}, where Vega served as a check star
(the Vega data in machine readable form are in a separate file in the
Supplementary Information).  The observations and much of the data
reductions are exactly as described for Altair\cite{Peterson06} which
should be consulted for details.  Issues specific to the Vega data are
described in the Supplementary Information.  We focus here on the
closure phase data taken by NPOI on May 25, 2001, as they are the most
extensive and of the highest quality data of that run.  We augment
these with the $V$-band magnitude\cite{Bohlin04}, $V=0.026\pm 0.008$,
as an additional observable.  The model was fitted by enforcing the
usual minimum $\chi^2$ metric using the Levenberg--Marquardt
algorithm\cite{Press92}.

The parameters from the initial reduction are given in column 2 of
Supplementary Table~S1.  The projected (at inclination, $i$) equatorial
velocity, $v_{\mathrm{eq}}$, for this model was predicted to be
$v_{\mathrm{eq}}\sin i\sim 15$\,\kms, a bit below the $\sim21.8 \pm
0.1$\,\kms\ found from detailed profile fits using a rotating
model\cite{Hill04}.  Although it is not so far off given how close the
star is to pole-on, we feel the projected velocity is sufficiently well
known that it should also be incorporated in the fit.  We have
therefore added as an ``observable'' $v_{\mathrm{eq}}
\sin i = 22\,\pm1.0$\,\kms (see the Supplementary Information) and
refit the data (columns 3 and 4 of Supplementary Table S1).

\begin{figure}
\includegraphics[angle=-90,scale=0.90]{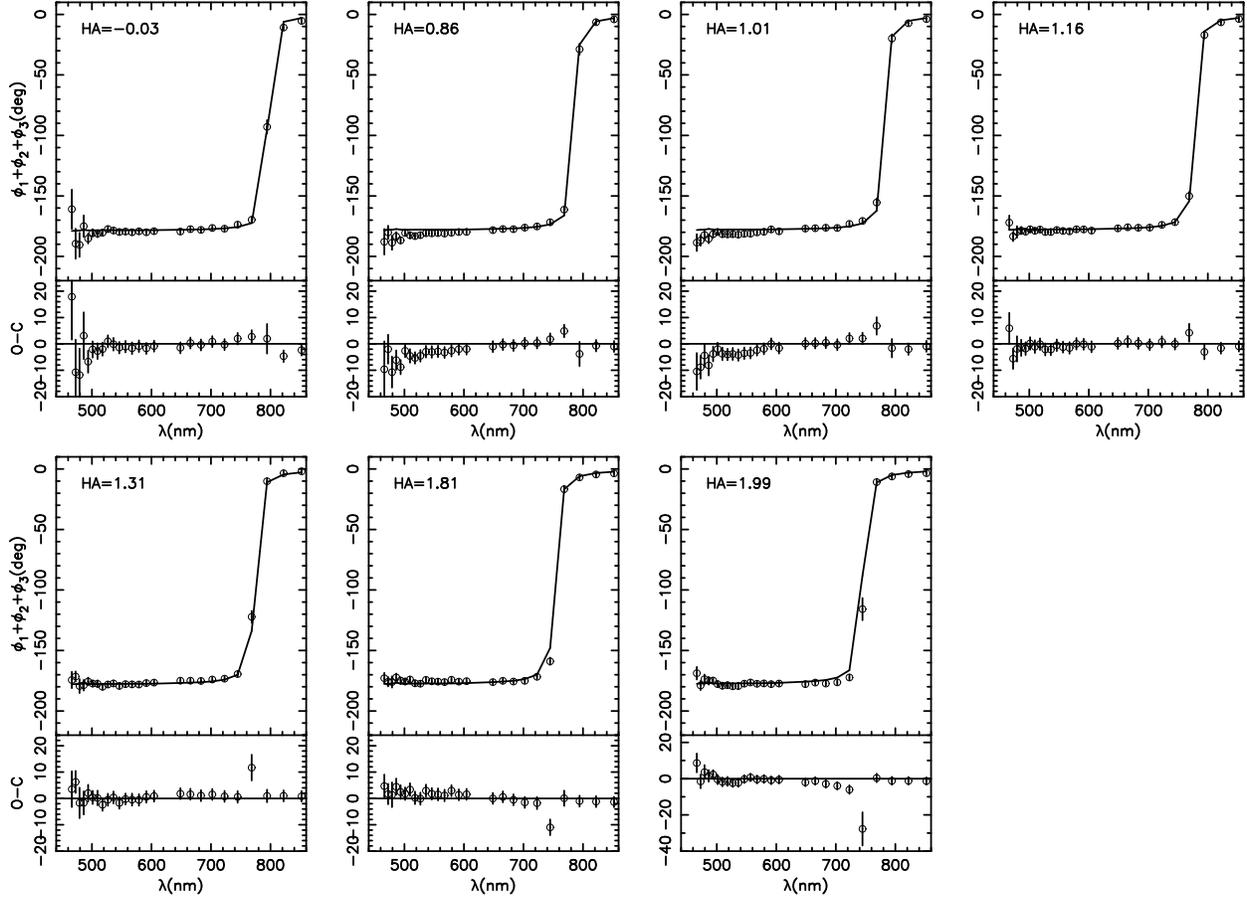}
\caption{{\bf Roche model fits to the closure phase data taken May 25,
2001.} The observations (open circles), estimated errors (bars,
standard deviations) and model calculations (solid lines) are shown for
each scan (labeled by hour angle, HA).  Residuals (observed, O, minus
calculated, C) are shown below each of the scans for clarity.  The
phases for the individual baselines, $\phi_i$, and thus the closure
phases, take on only two values, $0^{\circ}$ or $180^{\circ}$, if an
object is centro-symmetric.  Closure phase measurements showing
departures from this simple ``abrupt transition'' behavior provide
potentially very sensitive measurements of asymmetry in an object.  The
soft transition at the points of the 180$^{\circ}$ phase changes here
give a clear signal of the asymmetry in the intensity
distribution. (Note the scale change for the residuals of the last
scan.)\label{fig:fits}}
\end{figure}

\begin{figure}
\begin{center}
\includegraphics[angle=0,scale=0.7]{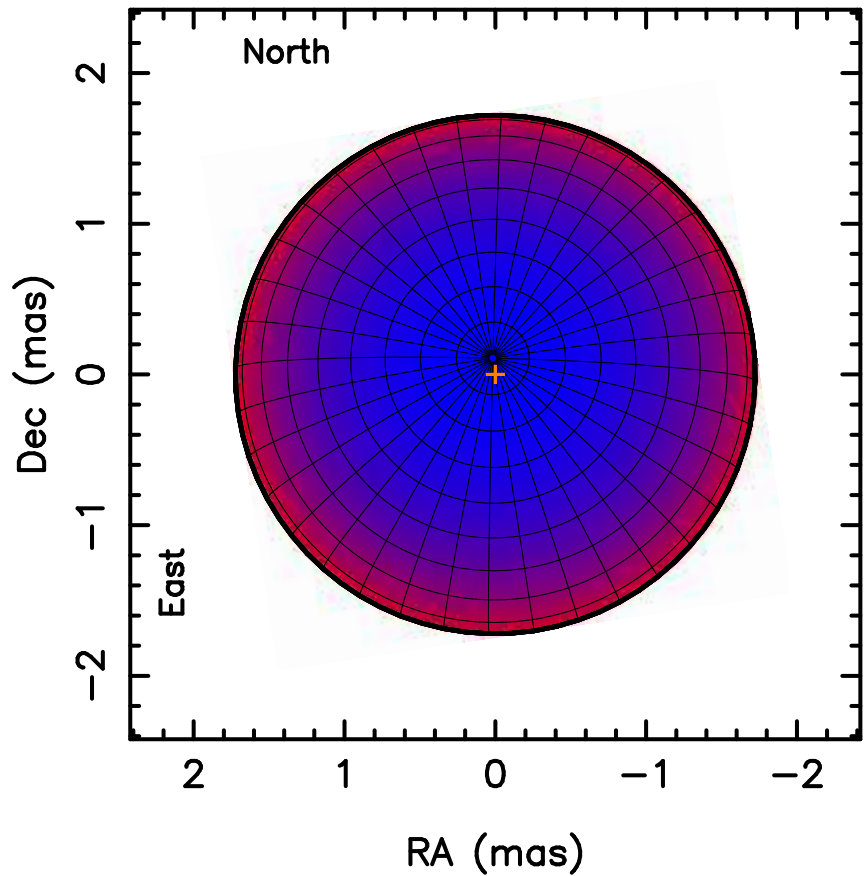}
\caption{{\bf A false color model of Vega as it appears from Earth.}
(Blue is bright, red is faint, and the orange ``+'' is the subsolar
point.)  The temperature drops more than 2400\,K from pole to equator,
creating an 18$\times$ drop in intensity at 500\,nm.  Limb-darkening in
a non-rotating model predicts only a 5-fold drop in intensity.
Although the projected outline is almost perfectly circular, the polar
diameter is only 80\% of the equator. Dec., declination; RA, right
ascension. \label{fig:model}}
\end{center}\end{figure}

The Roche model provides a good fit to the augmented data set.  The
model parameters fit, a number of derived quantities, and an estimate
of the main parameters of the star Vega would be, were it not rotating,
are given in Table~1.  The quality of the fit is illustrated in
Figure~\ref{fig:fits} and a false color rendering of the Vega model is
shown in Figure~\ref{fig:model}.  As Vega plays so many roles in
astronomy, the ramifications of this result are extensive.  We
summarize some of the most important below.

%Fig: Vega-obs

%Fig: Vega-model

Rotation can affect the gross spectral distribution of a star, an issue
of considerable import given Vega's role as the primary flux calibrator
in the ultraviolet, visible and near infrared.  To estimate the size of
these effects we have calculated the changes in the fluxes from the
rotating model compared to the static case that would be measured
through a series of broadband filters (Table~2).  As can be seen, there
is a significant, systematic increase in the infrared emission from the
rotating model.  This is understood as due to the large amount of
surface area predicted to be at relatively low temperatures.

There is an extensive literature on the possibility and extent of an
``infrared excess'' in the Vega spectrum\cite{Blackwell83,Leggett86b}.
As the issue is critical to so much of astronomy, sides have been
strongly taken. Observations of other A stars showed that Vega's colors
appear sensibly normal\cite{Leggett86b}, which led
authors\cite{Leggett85,Leggett86b} to wonder whether the problem was
with the model atmospheres.  Others\cite{Bess98} argued that because
the hydrogen absorption coefficient completely dominates the spectra of
A stars and is so well known, it was unlikely the model atmospheres
were wrong, an argument that seems to have carried the day.  Rapid
rotation provides a simple resolution to this controversy, Vega is best
modelled as a composite of model atmospheres.  The star should have
``normal'' infrared colors, as most normal A stars are rapid rotators,
and there is no need to question the quality of the individual
atmosphere models.

One important aspect to this model is the near pole-on orientation of
the rotational axis.  Vega is surrounded by a large infrared emitting
disk of material, a ``debris disk'', which presents an essentially
circular profile.  One does not expect perfect coupling between
orbiting material and the central star.  But, if the poles of the disk
and Vega coincide and the disk is thin we would predict 0.3\%
flattening, which is unlikely to be detectable, as seems to be the
case\cite{Su05}.

Related to the orientation of the pole is the inferred equatorial
velocity, $v_{\mathrm{eq}} = 272$\,\kms\ (Table~1).  Among the
Vega-like stars, Vega itself has been anomalous in displaying a very
low projected rotational velocity\cite{Song01}.  The present
determination strongly supports the view that the Vega-like stars are
rapid rotators, consistent with the large amounts of angular momentum
in the surrounding dust clouds.

Vega's rotational state also affects inferences about its debris disk.
It is well known that rotation results in the apparent brightness, and
in turn the deduced luminosity, being
inclination-dependent\cite{Collins63}.  Song \etal\cite{Song01}, for
example, have gone to some length to characterize the uncertainty
introduced by this effect in the Vega-like stars.  It is
straightforward to calculate the total luminosity of a Roche spheroid
(Table~1).  Further, one can apply small corrections\cite{Peterson06}
to the luminosity and polar radius to obtain the corresponding values
that would apply to a non-rotating star of the same mass.  Using the
Padova\cite{Girardi00} models we derive $M=2.303\,\Msun$ and an age of
$386\pm16$\,Myr, on the high side of recent estimates of
$354^{+29}_{-87}$\,Myr\cite{Song01} and
$347^{+43}_{-37}$\,Myr\cite{Habing99}.

Unfortunately, Vega's composition enters the age determination rather
critically.  The star is currently viewed as underabundant in heavy
elements compared to the Sun\cite{Castelli94}, [Fe/H]\ $\sim -0.5$,
which clearly needs to be revisited, given the 2400\,K temperature drop
now predicted across its surface.  Previous age determinations have
implicitly or explicitly assumed solar metallicity, the argument being
that composition peculiarities among the slowly rotating stars in this
part of the Hertzsprung-Russell diagram are probably limited to the
outer envelope, and normal composition models are therefore appropriate
for estimating bulk properties.  In our interpretation that argument
fails, because at these rotational velocities meridional circulation
will keep the the bulk of the star well mixed.

Proceeding as above we estimate $M=2.11\,\Msun$ and an age of 572\,Myr,
using the $Z=0.008$ ([Fe/H]\ $\sim-0.4)$ Padova models\cite{Girardi00}.
It is clear that a full abundance analysis incorporating rotation needs
to be performed to remove this uncertainty.  In the meantime this
range, 386--572\,Myr, is probably a more realistic estimate of the
uncertainty in the evolutionary age of Vega and its debris disk.

\noindent{\bf Received 10 January; accepted 14 February 2006.}